\documentclass[10pt,letterpaper]{article}
\usepackage{opex3}
\usepackage{color}
\usepackage{ulem}
\usepackage{graphicx}
\usepackage{caption}
\usepackage{subcaption}
\usepackage{soul}

\usepackage{upgreek}

\begin{document}


\title{Compensation of anisotropy effects in the generation of two-photon light}

\author{Andrea Cavanna$^1$, Angela M. P\'erez$^1$, Felix Just$^1$, Maria V. Chekhova$^{1,2,3}$ and Gerd Leuchs$^{1,3}$}

\address{$^1$ Max Planck Institute for the Science of Light, G\"unter-Scharowsky-Stra{\ss}e 1/Bldg 24, 91058 Erlangen, Germany\\
$^2$ Department of Physics, M. V. Lomonosov Moscow State University, Leninskie Gory, 119991 Moscow, Russia\\
$^3$ University of Erlangen-N\"urnberg, Staudtstrasse 7/B2, 91058 Erlangen, Germany }

\email{andrea.cavanna@mpl.mpg.de} 



\begin{abstract}
We analyse a method to compensate for anisotropy effects in the spatial distribution of parametric down-conversion (PDC) radiation in bulk crystals. In this method, a single nonlinear crystal is replaced by two consecutive crystals with opposite transverse walk-off directions. We implement a simple numerical model to calculate the spatial distribution of intensity and correlations, as well as the Schmidt mode structure, with an account for the anisotropy. Experimental results are presented which prove the validity of both the model and the method.
\end{abstract}


\section{Introduction}

Low-gain (spontaneous) parametric down-conversion (PDC) is one of the most common sources of entangled photons~\cite{burnham70}. At high parametric gain, PDC generates bright squeezed vacuum (BSV) where the photon number per mode can reach several orders of magnitude~\cite{timur12,kirill12}. In both high-gain and low-gain regimes, PDC is an essentially multimode source and a lot of effort is spent to prepare its output  in a single-mode state \cite{mosley08,branczyk10,straupe11,dixon13} since many applications require single-mode beams. In order to address the problem one has to understand the characteristics of these modes. This is usually done by employing a decomposition of the state into its coherent modes, called the Schmidt  decomposition. When the patterns of the modes are known, it is possible to employ some kind of filtering that matches the shape of a single mode.

In many cases, the spatial distribution of PDC radiation can be well described in the double-Gauss approximation~\cite{fedorov06}. Then, the Schmidt decomposition can be performed analytically and the modes are Hermite-Gaussian~\cite{straupe11, walborn12} (alternatively, Laguerre-Gaussian basis can be used). However, to achieve the high-gain regime, one should focus the pump tightly and use longer crystals, and then the effect of transverse walk-off becomes noticeable. It manifests itself in the asymmetry of the spatial distributions for both the intensity and the correlation function~\cite{fedorov07}. The resulting angular spectrum can no longer be decomposed into Hermite-Gaussian or Laguerre-Gaussian modes.

It has been known for decades that high-gain parametric amplification can be improved by using two crystals with the optic axes tilted oppositely~\cite{bosenberg89,armstrong97} and this configuration was used in the first experiments on squeezing~\cite{slusher87}. At the same time, no analysis of the spectral distribution was made and the Schmidt decomposition was never discussed.  In this paper we study the angular structure of the PDC radiation generated in the walk-off compensating configuration. We will restrict ourselves to the case of two-photon light since for this regime the theoretical model is well known and we can compare the experimental results with the calculation. A similar experiment with BSV has been described in Ref.~\cite{perez13}.

\section{Theory}
The quantum state of two-photon light generated via PDC is given by
\begin{equation}
\label{eqn:biphotonState}
| \psi \rangle = \int \int \mathrm{d}\theta_s \mathrm{d}\theta_i F(\theta_s,\theta_i) \hat{a}^{\dagger}_s \hat{a}^{\dagger}_i |0\rangle,
\end{equation}
where $F(\theta_s,\theta_i)$, known as the two-photon amplitude (TPA), is the probability amplitude that the signal photon is emitted at an angle $\theta_s$ and the idler photon at an angle $\theta_i$. Alternatively, the TPA can be written in the transverse coordinate representation, as the probability amplitude $F(x_s,x_i)$ that the signal photon leaves the crystal at the transverse position $x_s$ and the idler photon at the transverse position $x_i$.
Using this distribution it is possible to fully describe the spatial distribution of two-photon light, in particular to find the Schmidt decomposition and the shape of each mode.

The TPA can be found from the Hamiltonian $H$ of the PDC, as the state (\ref{eqn:biphotonState}) can be written in the first order of the perturbation theory,
\begin{equation}
|\psi\rangle=\exp \left(\frac{1}{i\hbar}\int dt H\right)|0\rangle.
\label{eqn:wavefunction}
\end{equation}

The Hamiltonian that describes the PDC process, in its turn, is proportional to the volume integral of the product of the three fields involved, namely the pump, the signal, and the idler,
\begin{equation}
\label{eqn:H}
\hat{H}\propto \int \mathrm{d}V \chi^{(2)} \left(E_p^{(+)}E_s^{(-)}E_i^{(-)}+\mathrm{h.c.}\right),
\label{eqn:TPA}
\end{equation}
where $V$ is the volume of the crystal in which the interaction takes place and $\chi^{(2)}$ is the quadratic susceptibility. The signal and idler fields $E_{s,i}$ must be described as quantum ones but since the pump $E_p$ is a strong laser field it is sufficient to consider it as a classical Gaussian beam.

\begin{figure}[htb]
\centering
\begin{subfigure}[b]{0.49\textwidth}
\includegraphics[width=\textwidth]{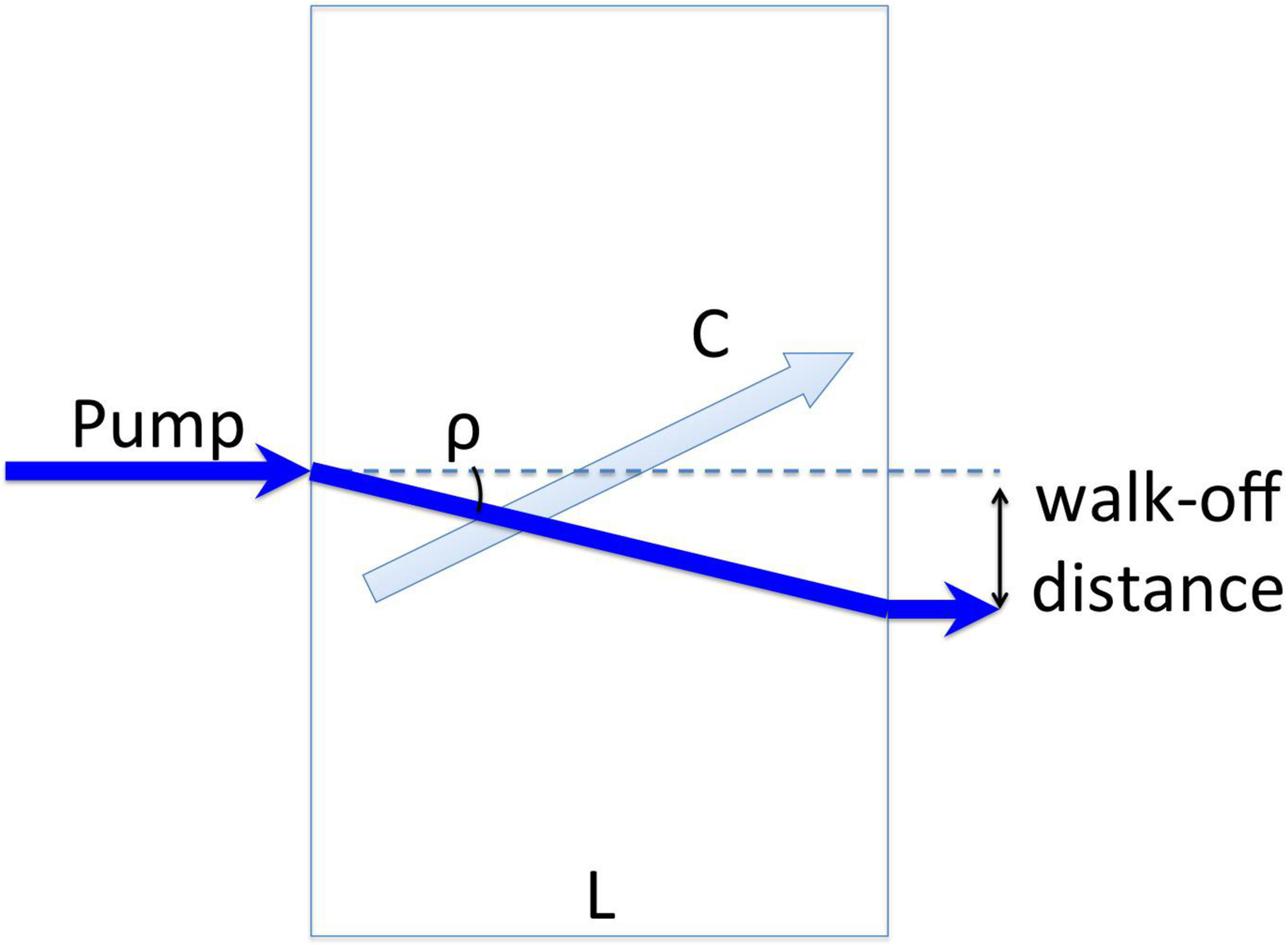}
\caption{}
\label{fig:oneCrystalAnisotropy}
\end{subfigure}
\begin{subfigure}[b]{0.49\textwidth}
\includegraphics[width=\textwidth]{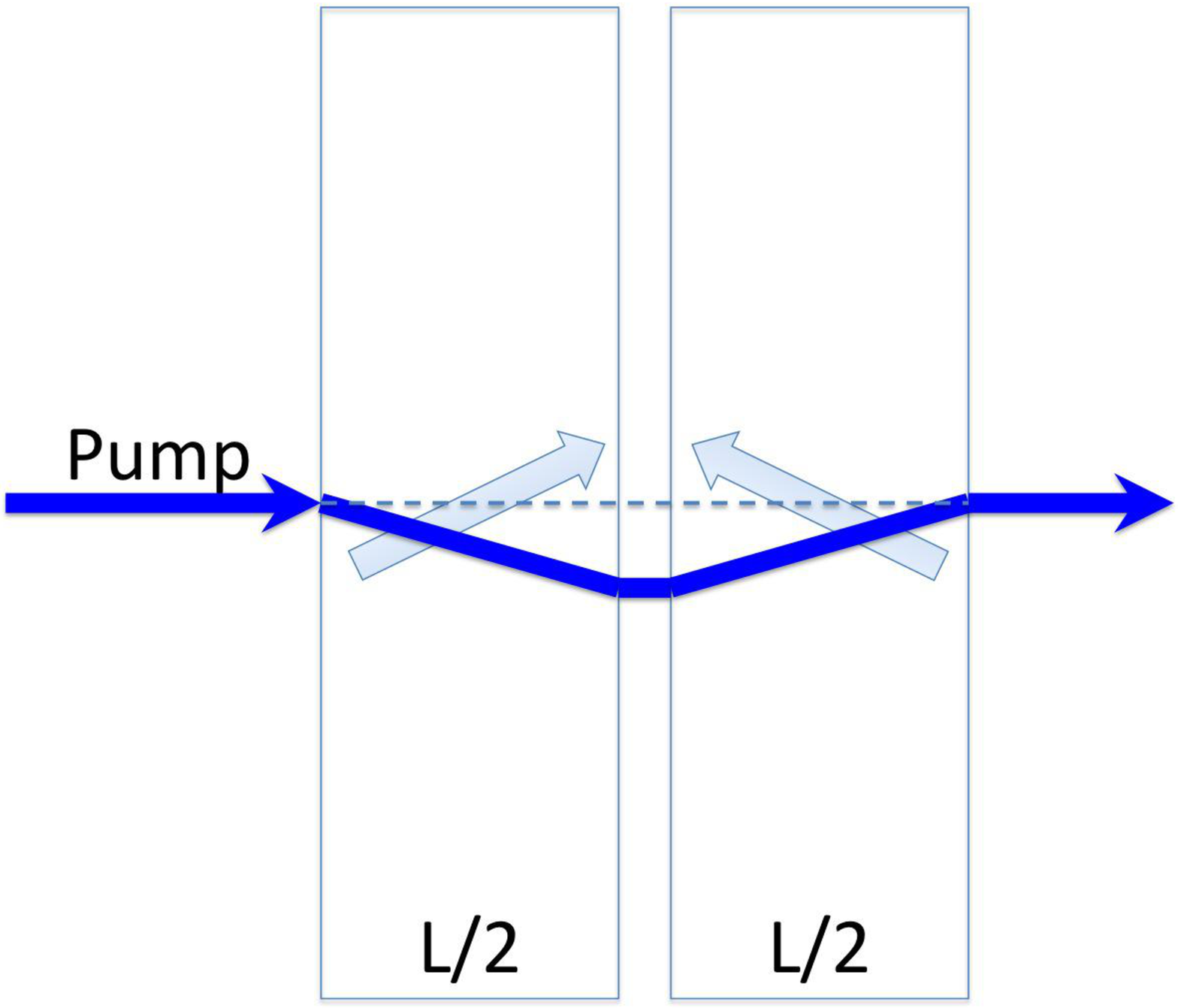}
\caption{}
\label{fig:anisotropyCompensation}
\end{subfigure}
\caption{(a) Transverse walk-off effect: an extraordinarily polarised beam impinging on a crystal normally is tilted while inside. The walk-off angle $\rho$ depends on the angle of the pump wave vector with the optic axis C and the crystal birefringence. (b) Compensation of the anisotropy effect is obtained by using two crystals with the optic axes tilted in opposite directions. In the two crystals, the walk-off, as well as the anisotropy, manifests itself in the opposite directions, so the effects cancel each other.}
\end{figure}

Because the pump beam usually has extraordinary polarisation, inside the nonlinear crystal it undergoes the walk-off effect. Namely, even if the pump is incident perpendicular to the crystal, its Poynting vector inside the crystal will be tilted by an angle $\rho$ (Fig.~\ref{fig:oneCrystalAnisotropy}), while the wave vector remains in the same direction as outside of the crystal. The direction of the walk-off is determined by the angle between the wave vector of the pump and the optic axis. When the pump waist is smaller or comparable to the transverse displacement due to the walk-off (the walk-off distance), the spatial distribution of PDC radiation is not symmetric any more~\cite{fedorov07}.

In order to compensate for the effect of anisotropy we use two crystals with opposite walk-off directions, as shown in Fig. \ref{fig:anisotropyCompensation}. The resulting spatial distribution is symmetric but due to the free space propagation between the crystals the PDC radiation generated in different crystals acquires a relative phase. The interference between light generated in the first and in the second crystal leads to a fringe pattern. 

\begin{figure}
\centering
\begin{subfigure}[b]{0.49\textwidth}
\includegraphics[width=\textwidth]{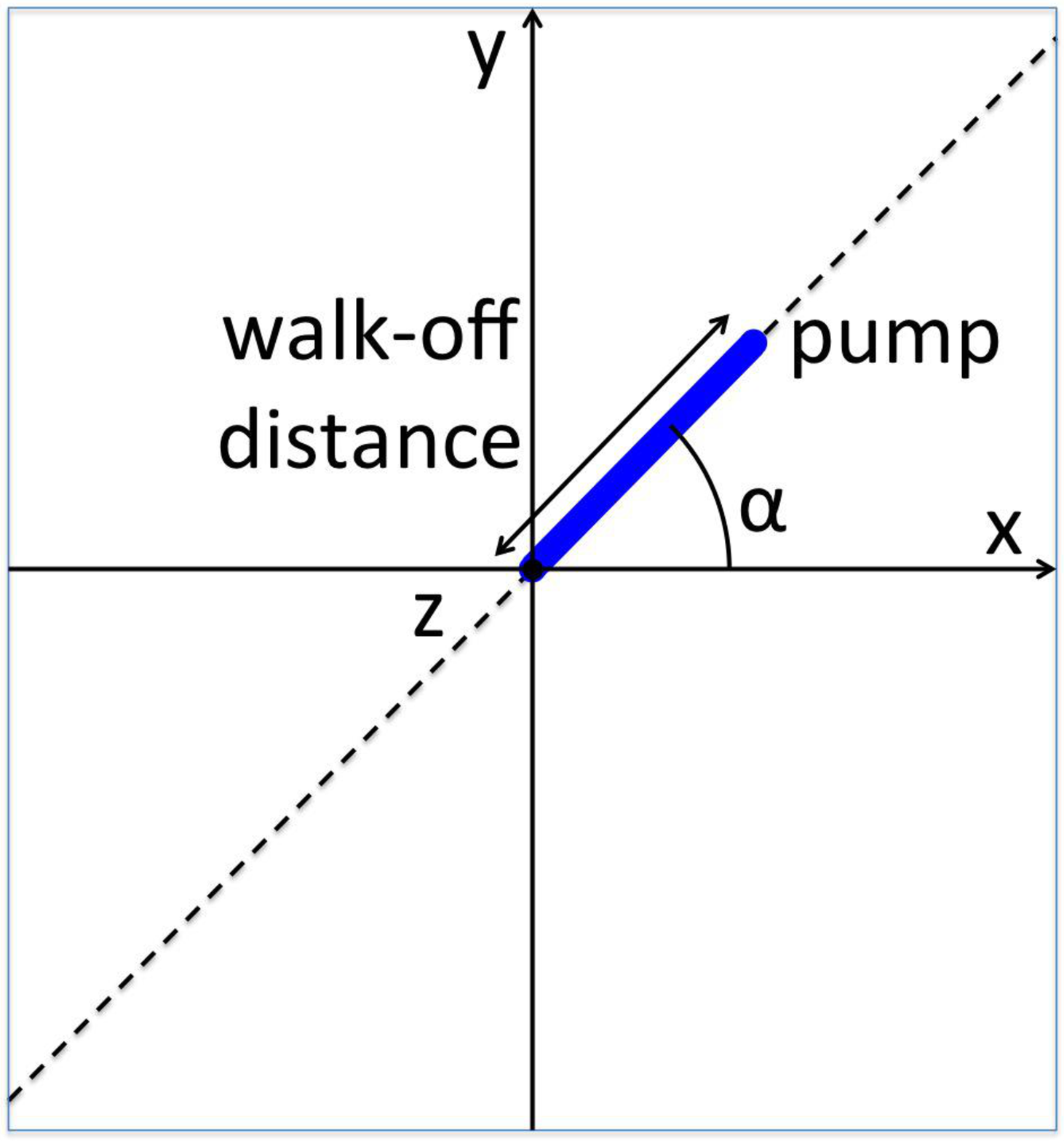}
\caption{}
\end{subfigure}
\begin{subfigure}[b]{0.49\textwidth}
\includegraphics[width=\textwidth]{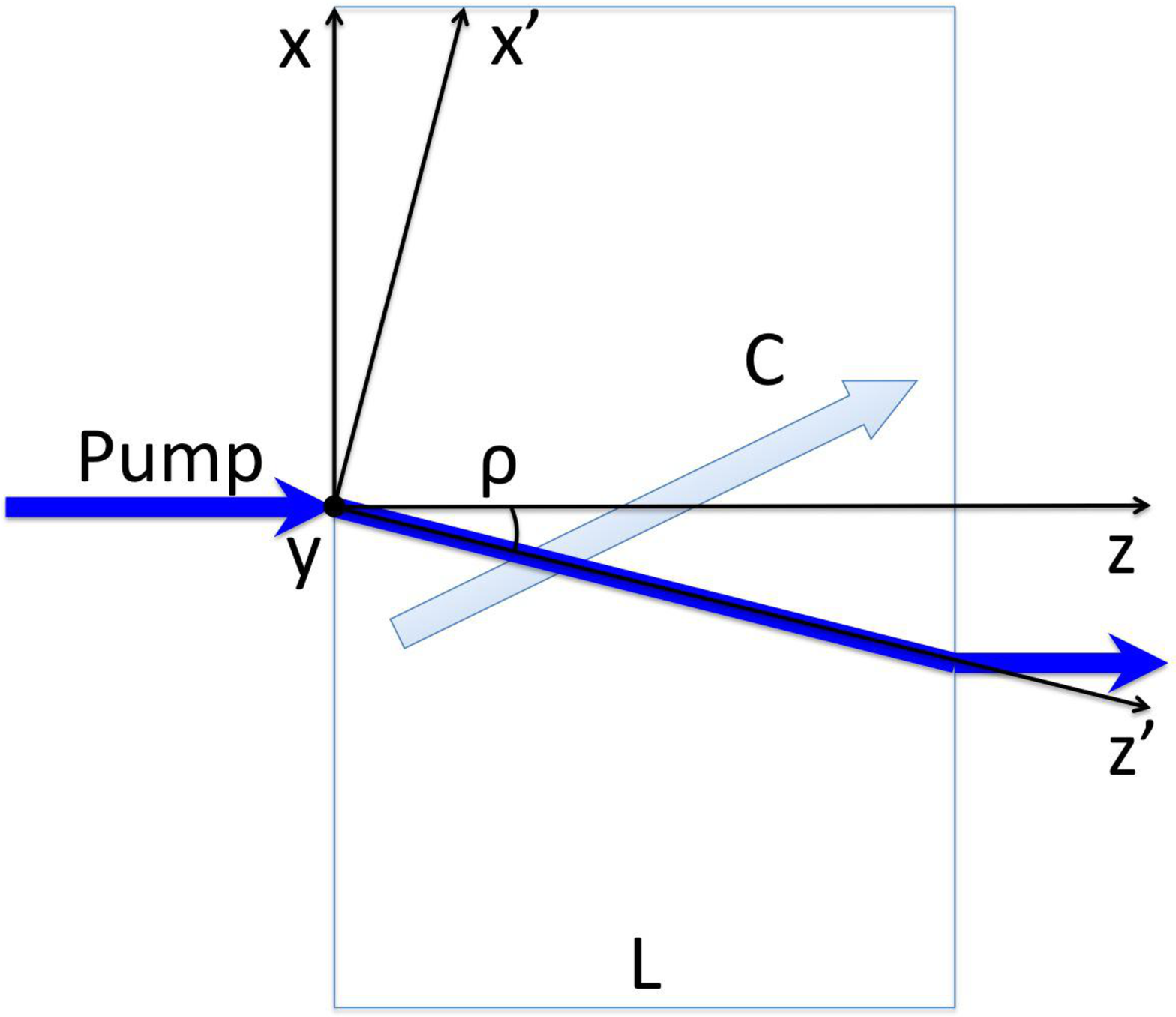}
\caption{}
\end{subfigure}
\caption{(a) Front view of the crystal with the principal plane (shown by dashed line) tilted by an angle $\alpha$ with respect to the plane $x-z$ in which the angular spectra are considered. (b) Tilted frame of reference in the principal plane for the case $\alpha=0$.}
\label{fig:geometry}
\end{figure}

In our consideration, the system consists of two BBO crystals cut for collinear type I ``eoo" phase matching such that the pump is impinging perpendicular to their entrance faces. For this configuration there are no additional effects due to refraction. The anisotropy manifests itself in the asymmetry of the angular distributions of the intensity and two-photon correlations. It is mainly visible for the angular distributions in the principal plane, that is, in the plane formed by the pump wavevector and the optic axis. However, as we show further, it also reveals itself in the plane orthogonal to the principal one, although in this case the effect is more subtle. Further, we will refer to these two cases as the ones of `strong anisotropy' and `weak anisotropy'. Accordingly, we will assume the principal plane to be oriented arbitrarily with respect to the `laboratory' axes $x,y$, see Fig.~\ref{fig:geometry}. The angle between the principal plane and the $x$ axis will be denoted by $\alpha$. The $z$ axis of our frame of reference coincides with the direction of the pump outside the crystal. The polarisation of the pump is always considered parallel to the principal plane, in order to provide phase matching. Correspondingly, inside the crystal, the pump beam propagates at the walk-off angle $\rho$ to the $z$ axis (Fig.~\ref{fig:geometry}).
In our measurement, the scanning of the angular distributions was always along the $x$ axis, the `strong anisotropy case' corresponded to $\alpha=0^\circ$ and the `weak anisotropy case', to $\alpha=\pi/2$.

In order to calculate the TPA, one should perform the volume integration in (\ref{eqn:H}) and then substitute the Hamiltonian in Eq. (\ref{eqn:wavefunction}). The fields are represented, as usual, through the photon creation and annihilation operators in plane-wave modes~\cite{klyshko88} and for the pump field we assume a gaussian distribution propagating along the path shown in Fig.~\ref{fig:geometry}. Similar to the way it was performed in Ref.~\cite{perez13}, we do the integration by passing to a new frame of reference, but in a more general 3D case:
\begin{eqnarray}
x' &=& \left[\left(y\sin{\alpha}+x\cos{\alpha}\right)\cos{\rho}-z\sin{\rho}\right]\cos{\alpha}-\left(y\cos{\alpha}-x\sin{\alpha}\right)\sin{\alpha} \nonumber \\
y' &=& \left(y\cos{\alpha}-x\sin{\alpha}\right)\cos{\alpha}+\left[\left(y\sin{\alpha}+x\cos{\alpha}\right)\cos{\rho}-z\sin{\rho}\right]\sin{\alpha} \\
z' &=& \left(y\sin{\alpha}+x\cos{\alpha}\right)\sin{\rho}+z\cos{\rho}. \nonumber
\label{eqn:newvariables}
\end{eqnarray}

The TPA is then calculated as
\begin{equation}
F(\theta_s,\theta_i)=\int\int dx'dy'\int_{-L/2\cos\rho}^{L/2\cos\rho} dz'\,\exp\left(-\frac{x'^2+y'^2}{2\sigma^2}\right)\exp\left(i\Delta k_xx+i\Delta k_yy+i\Delta k_zz\right),
\label{eq:generalPDC}
\end{equation}
where $\Delta k_{x,y,z}$ are the projections of the wave-vector mismatch, $\Delta \mathbf{k} = \mathbf{k}_p -\mathbf{k}_s - \mathbf{k}_i$, on different axes and $\sigma$ characterises the width of the pump field.

For our simulations we performed a numerical integration of the function $F(\theta_s,\theta_i)$. For the comparison with the experiment, two one-dimensional distributions of the TPA are calculated.
If one projects the TPA onto one of the axes ($\theta_s$ or $\theta_i$) the so-called unconditional distribution is obtained. This correspond to measuring the angular distribution of the signal or the idler radiation separately. Furthermore, the slice through the TPA at a fixed value of $\theta_i$, in other words the probability to observe a signal photon conditioned on the event of having the idler photon at $\theta_i$,  is known as the conditional distribution.

Until now only one crystal was taken into account. If a second crystal, identical to the first one, is positioned right after it, the TPAs of the two crystals will add up.
If both crystals have the same orientation of the optical axis, then also the anisotropy will be equal in both crystals. We will refer to this situation as the non-compensation case. When the second crystal is rotated by $180^\circ$ along the $x$ axis (Fig. \ref{fig:anisotropyCompensation}) then the effects of anisotropy in the two crystals are opposite.
The total probability distribution of the radiation generated by both crystals is given by the sum of their TPAs with the relative phase~\cite{zondy94} that depends on the distance $d$ between the two crystals,
\begin{equation}
F(\theta_s,\theta_i) = F_1(\theta_s,\theta_i)\cdot \exp\left[i\phi\right] + F_2(\theta_s,\theta_i),
\label{eqn:phase}
\end{equation}
where $\phi = d (k_{p0} -k_{s0} - k_{i0})$, $k_{p0},\,k_{s0}$, and $k_{i0}$ are the wave vectors of the pump, signal, and idler radiation in the space between the crystals and $F_1$, $F_2$ are the TPAs of the first and the second crystal respectively. These differ in three parameters: the limits of integration for $z'$ that are from 0 to $L/(2\cos\rho)$ for the first crystal and from $L/(2\cos\rho)$ to $L/\cos\rho$ for the second, the sign of $\rho$ that is opposite for the compensation configuration and the sign of $\chi^{(2)}$ that depends on the direction of the optic axis (shown by arrows in Fig.~\ref{fig:geometry}). Due to the phase factor the resulting TPA will show interference that can be constructive or destructive depending on the sign of $\chi^{(2)}$~\cite{burlakov97}.

\section{Experiment}

\begin{figure}
\centering\includegraphics[width=0.75\textwidth]{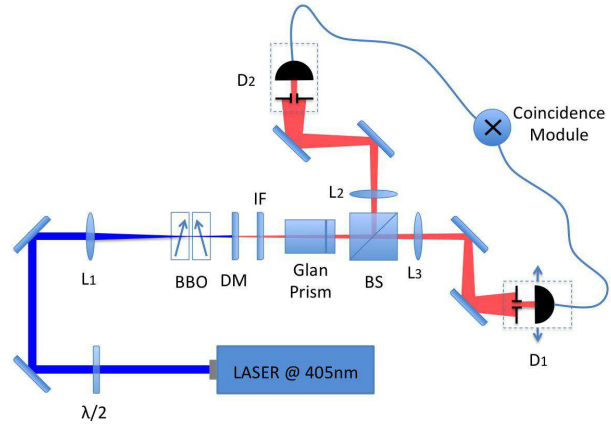}
\caption{The experimental setup: L$_1$ is a lens with $f=100$ mm, L$_2$ and L$_3$ have $f=200$ mm, DM is a dichroic mirror, IF is the interference filter and D$_1$ and D$_2$ are modules containing the detectors.}
\label{fig:setup}
\end{figure}

The experimental setup is shown in Fig. \ref{fig:setup}. In order to produce PDC we utilised two BBO crystals with $1$ mm length. The crystals were cut for collinear frequency-degenerate type-I phase matching, so that the optic axes formed an angle of $29.03^{\circ}$ with the $z$ axis. The pump laser operated in the continuous-wave mode at $405$ nm. A half wave plate (HWP) was used to change the polarisation of the pump to horizontal (which corresponded to extraordinary polarisation in our crystals). The two crystals were separated by a distance $d=5$ mm. Considering the parameters of the crystals and the pump, the walk-off angle was $3.85^\circ$ and hence the walk-off distance in one crystal was 67.4 $\upmu$m. In order to be able to observe the effect of anisotropy the beam waist should be at least comparable to the walk-off. This was achieved by focusing the pump with a lens ($f=10$ cm), with the beam waist right between the crystals. With the help of a CCD camera we have measured a beam waist of 78 $\upmu$m. In order to completely remove the pump radiation we used a dichroic mirror reflecting wavelengths below $450$ nm and a bandpass filter transmitting around 808 nm with 3 nm FWHM. A Glan-Thompson prism was further used to remove any remaining pump radiation. To separate the signal and idler radiation, we used a 50:50 beamsplitter. Each part of the split beam passed trough a lens with a focal length of $20$ cm. In the corresponding focal planes, two 100$\upmu$m pinholes were positioned. After that, the radiation was collected by fibre collimators and coupled to the single photon detectors (Si-APDs). In both arms the pinhole, together with the fibre coupler, was mounted on a translation stage. Thus it was possible to scan the position in order to measure the angular (far-field) distribution of the TPA. The relation between the $x$ coordinate and the angle $\theta$ is $x_{s,i}=f\tan{\theta_{s,i}}$, where $f$ is the focal lens of L$_2$ and L$_3$. The unconditional distribution is the spatial intensity distribution of idler or signal radiation and the conditional one is given by the coincidence distribution in the case where one detector is fixed while the other one is being scanned.
To measure in the weak anisotropy configuration, the polarisation of the pump was rotated by $90^{\circ}$, as well as both crystals and the Glan-Thompson prism. In this way the walk-off is in the vertical direction while the detectors scan in the horizontal plane.

\begin{figure}[htb]
\centering
\begin{subfigure}[b]{0.33\textwidth}
\includegraphics[width=\textwidth]{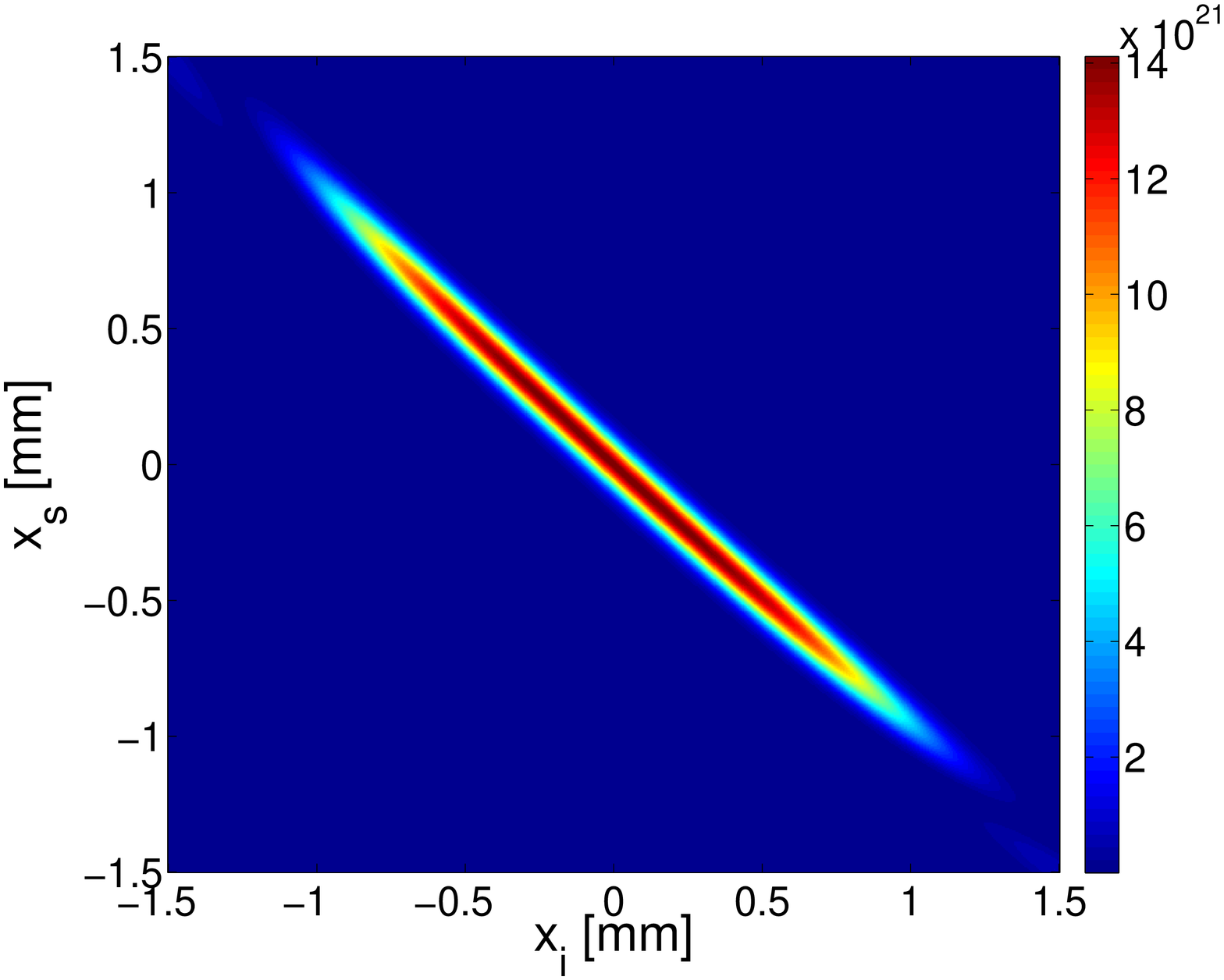}
\caption{TPA}
\label{fig:TPAoneCrys}
\end{subfigure}
\begin{subfigure}[b]{0.33\textwidth}
\centering
\includegraphics[width=0.8\textwidth]{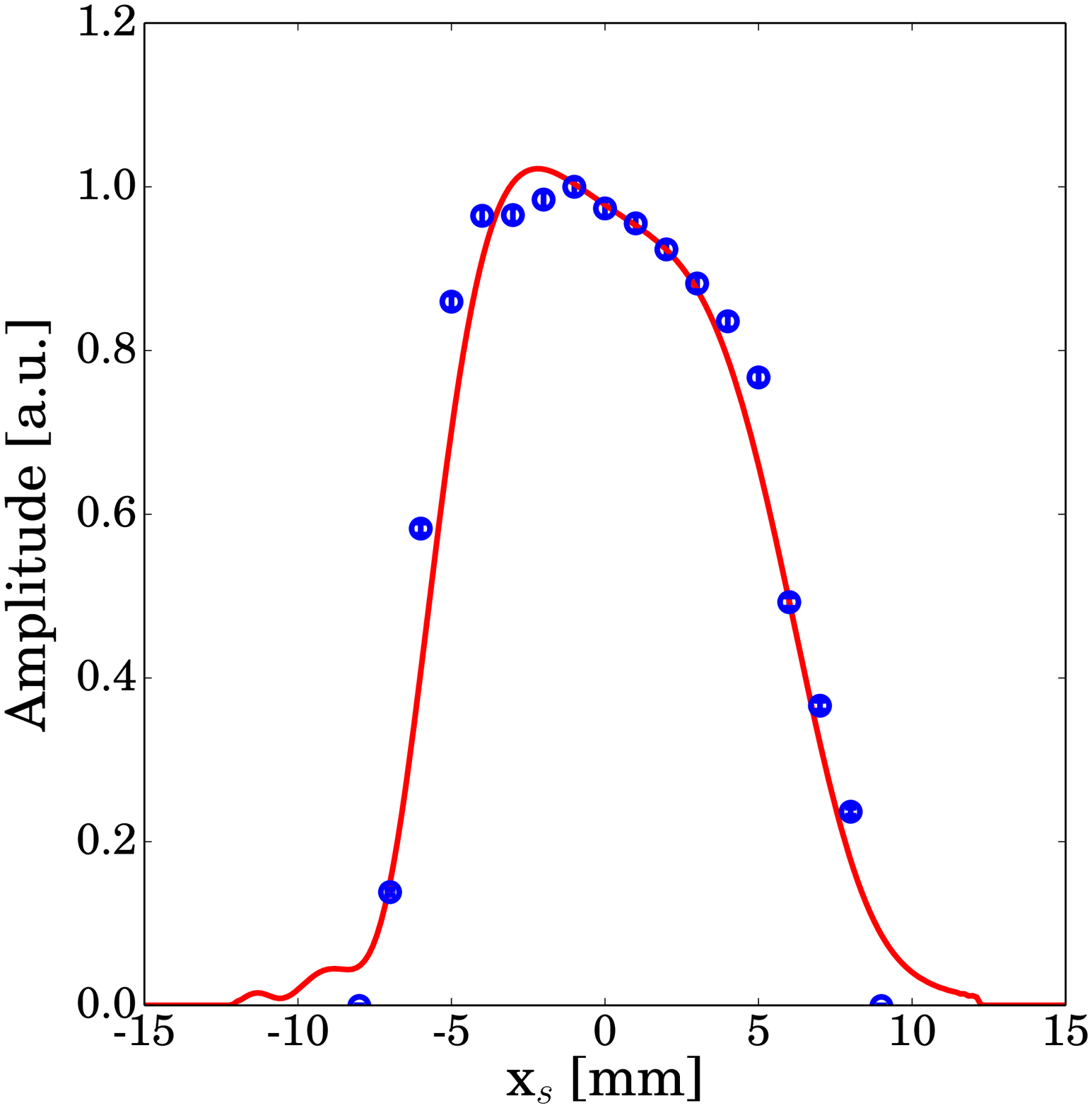}
\caption{Unconditional distribution}
\label{fig:UncondOneCrys}
\end{subfigure}
\caption{(a) TPA in the far field generated by $1$ mm BBO crystal with type-I collinear phase-matching. (b) The measured intensity distribution (blue points) and the theoretical prediction (red line).}
\label{fig:OneCrystalAnisotropy}
\end{figure}
First we have measured the transverse angular distribution of PDC generated by one crystal of $1$ mm length. The simulation and experimental results are shown in Fig. \ref{fig:OneCrystalAnisotropy}. The distortions introduced by anisotropy are evident. This is, first of all, the `bent' shape of the two-dimensional TPA distribution (Fig.~\ref{fig:TPAoneCrys}) and the resulting asymmetric unconditional distribution (Fig.~\ref{fig:UncondOneCrys}).

\begin{figure}[htb]
\centering
\begin{subfigure}[b]{0.32\textwidth}
\includegraphics[width=\textwidth]{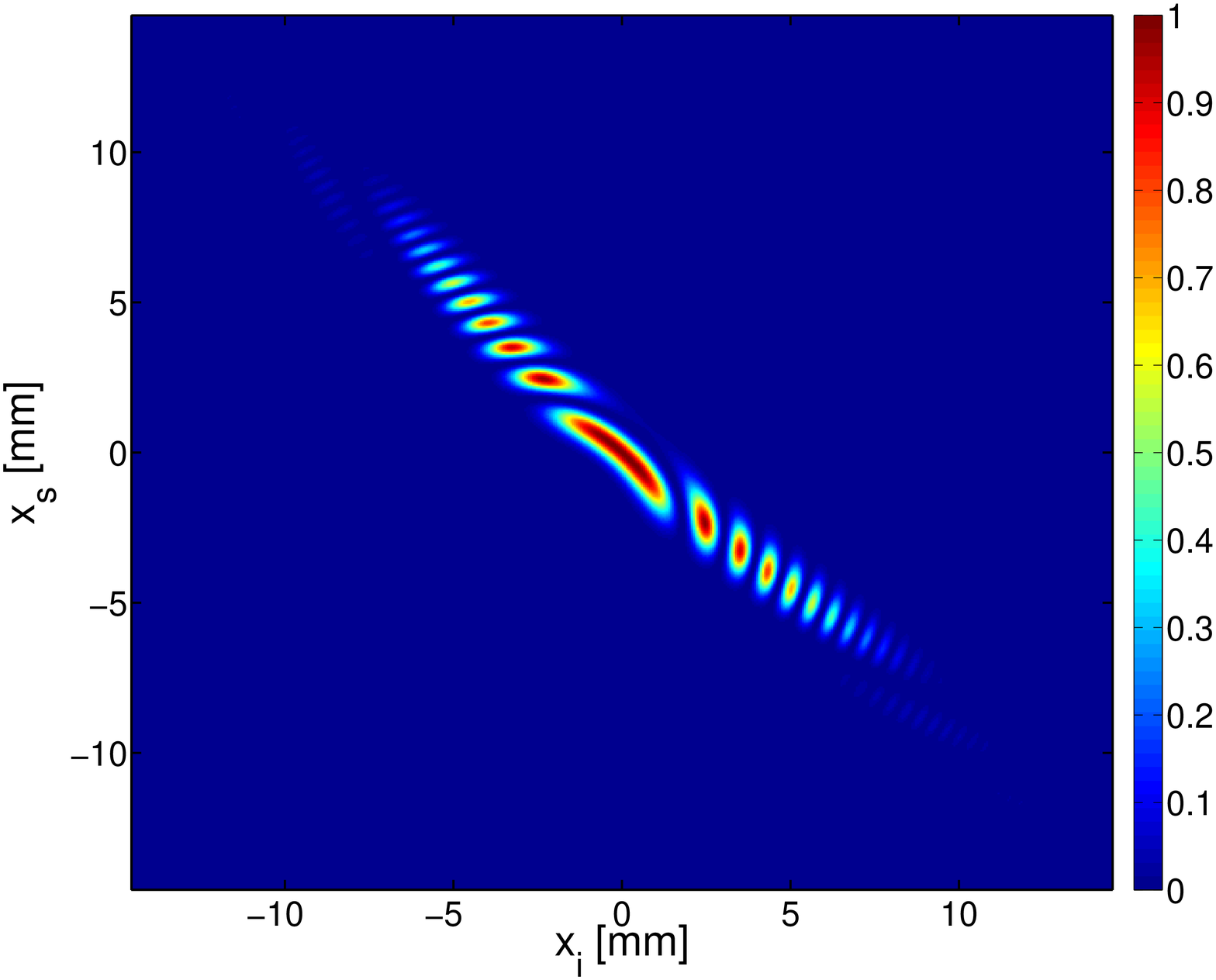}
\caption{TPA}
\label{fig:TPANoCompensation}
\end{subfigure}
\begin{subfigure}[b]{0.32\textwidth}
\includegraphics[width=\textwidth]{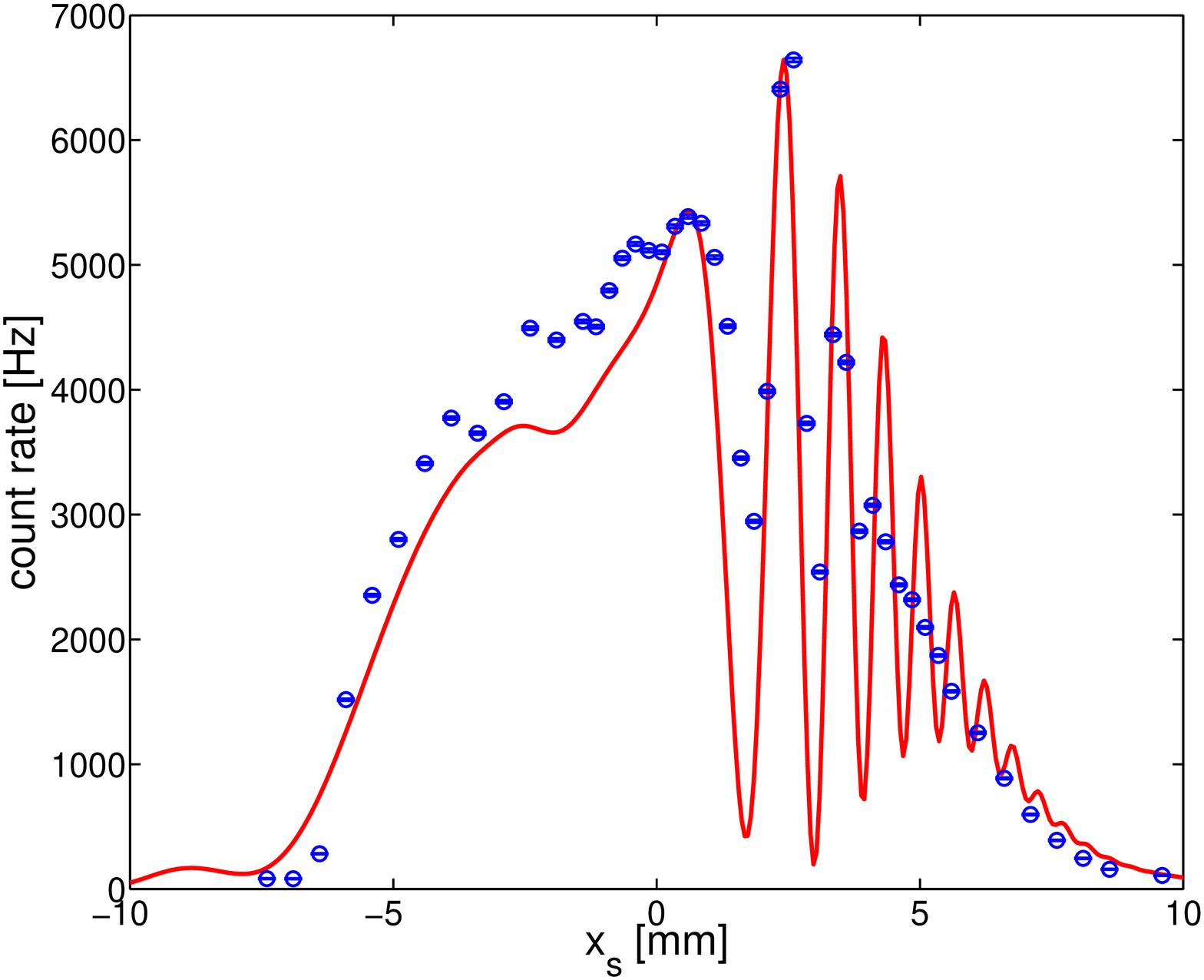}
\caption{Unconditional distribution}
\label{fig:UnconditionalNoCompensetion}
\end{subfigure}
\begin{subfigure}[b]{0.32\textwidth}
\includegraphics[width=\textwidth]{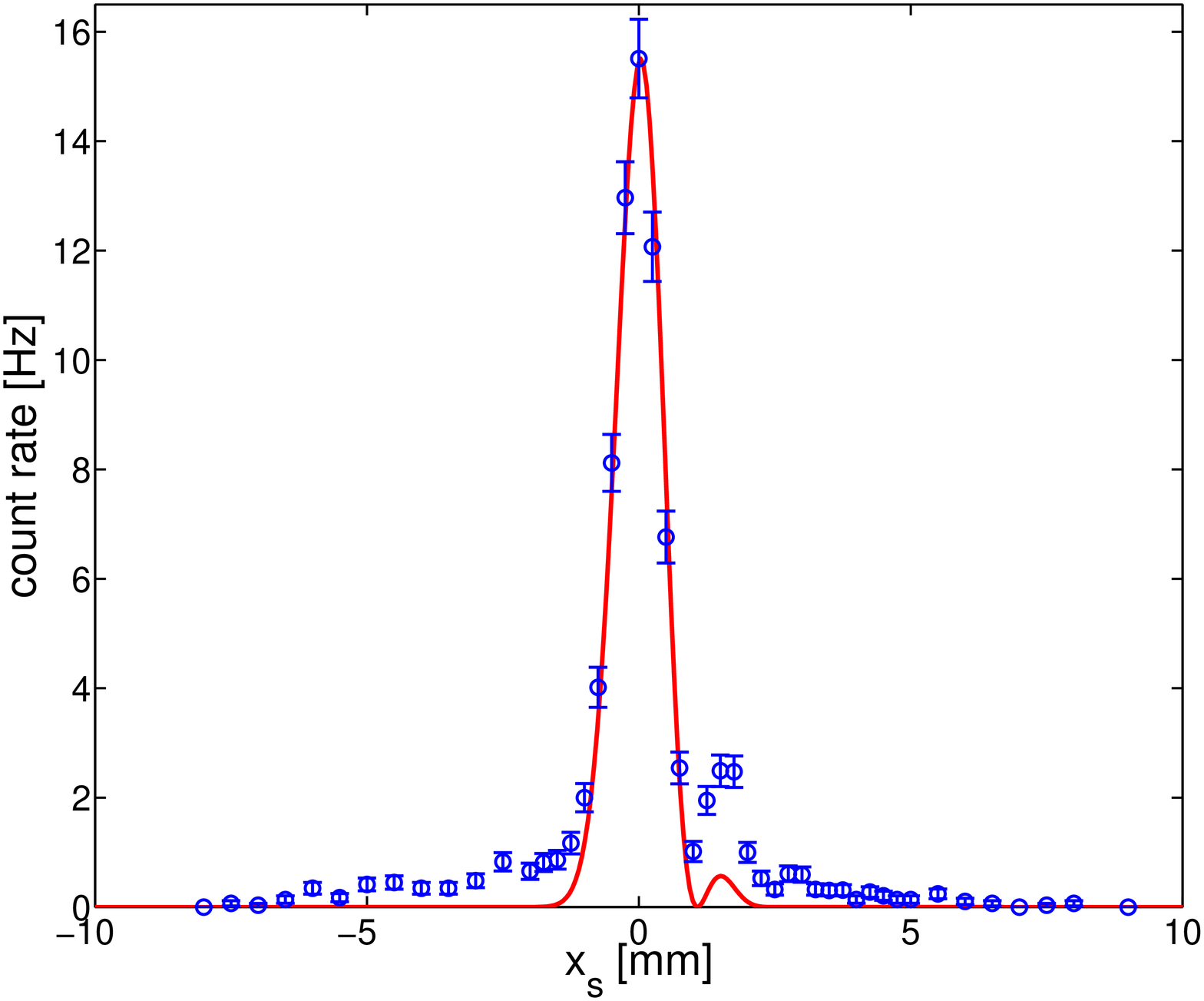}
\caption{Conditional distribution}
\label{fig:ConditionalNoCompensetion}
\end{subfigure}
\caption{Theoretical and experimental TPA distributions in the non-compensation configuration.}
\label{fig:twoCrysNoCompensation}
\end{figure}

Afterwards we included the second crystal, with its optic axis parallel to that of the first crystal (the non-compensation case). Figures~\ref{fig:UnconditionalNoCompensetion}, \ref{fig:ConditionalNoCompensetion} show the measured unconditional and conditional distributions as well as the theoretical predictions. As previously discussed, due to the space between the two crystals, the effect of interference is also visible.The reduction of visibility in the experimental data is mainly due to the width of the pinhole used to scan the distribution. The position of the fringes depends on the distance between the crystals. As it could not be measured with enough precision, it was chosen as a fitting parameter.
From Fig. \ref{fig:TPANoCompensation}, it is clear that the TPA becomes even more bent than in the single-crystal case. Accordingly, the unconditional distribution becomes even more asymmetric.

\begin{figure}[htb]
\centering
\begin{subfigure}[b]{0.32\textwidth}
\includegraphics[width=\textwidth]{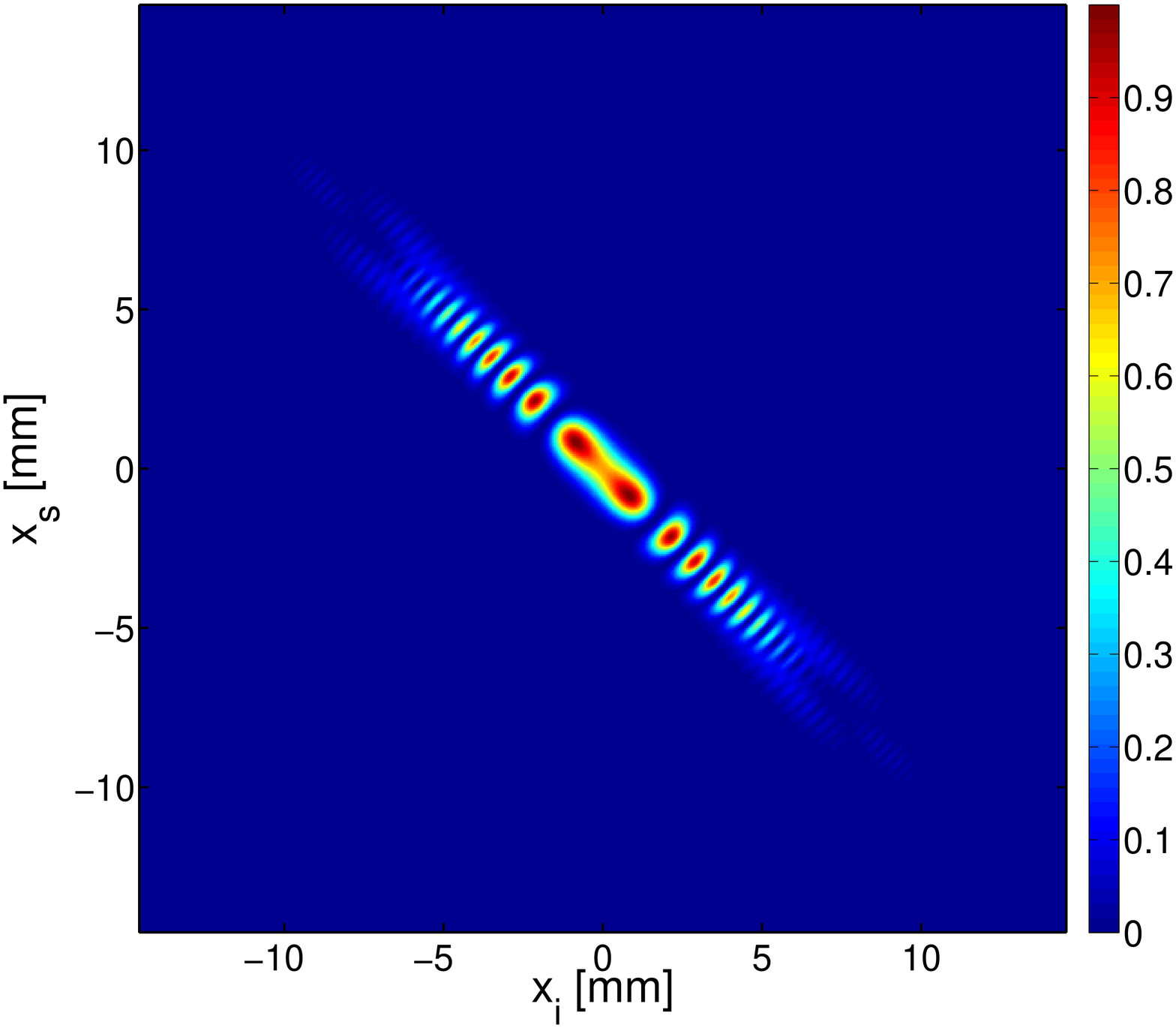}
\caption{TPA}
\label{fig:TPACompensation}
\end{subfigure}
\begin{subfigure}[b]{0.32\textwidth}
\includegraphics[width=\textwidth]{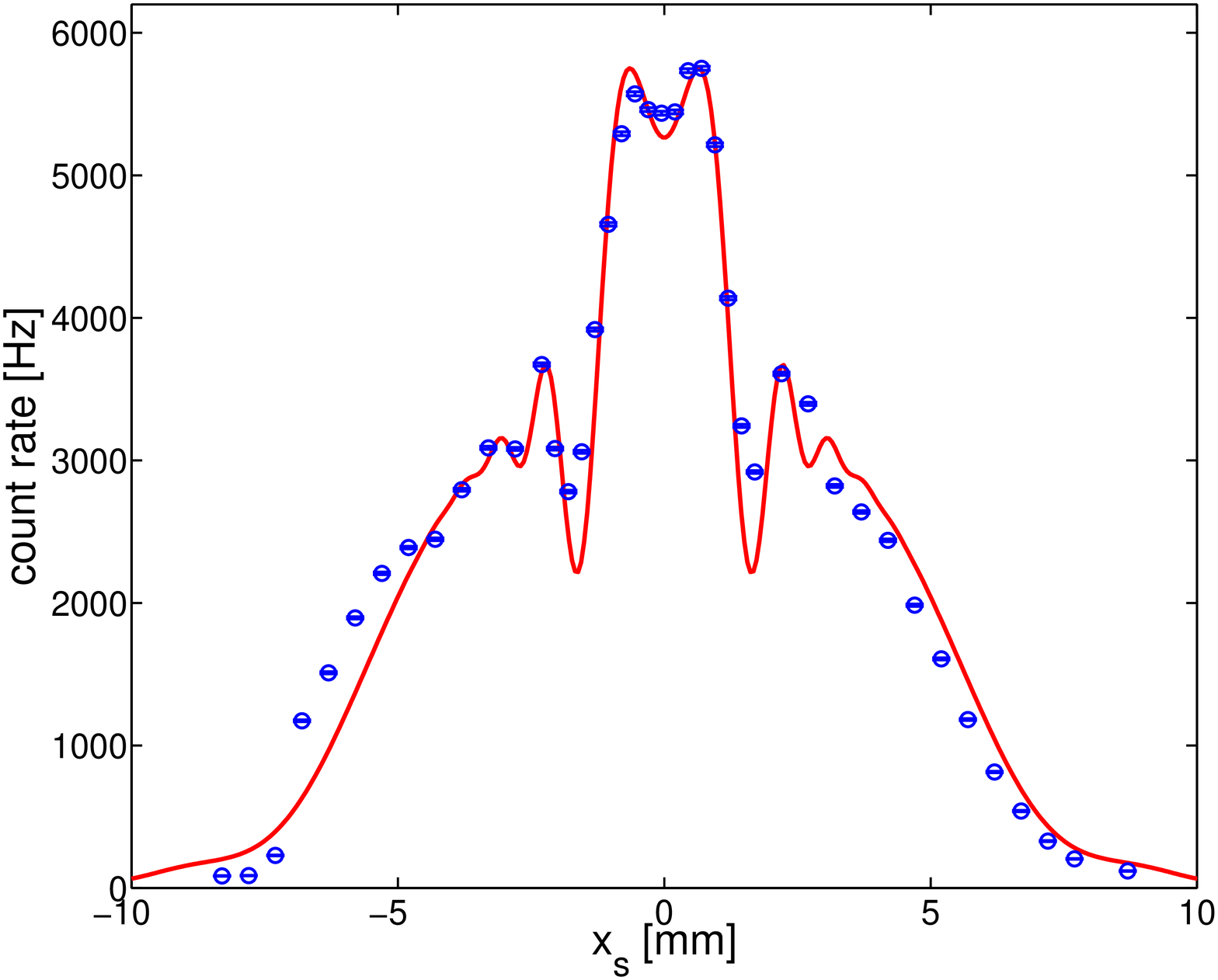}
\caption{Unconditional distribution}
\label{fig:UnconditionalCompensetion}
\end{subfigure}
\begin{subfigure}[b]{0.32\textwidth}
\includegraphics[width=\textwidth]{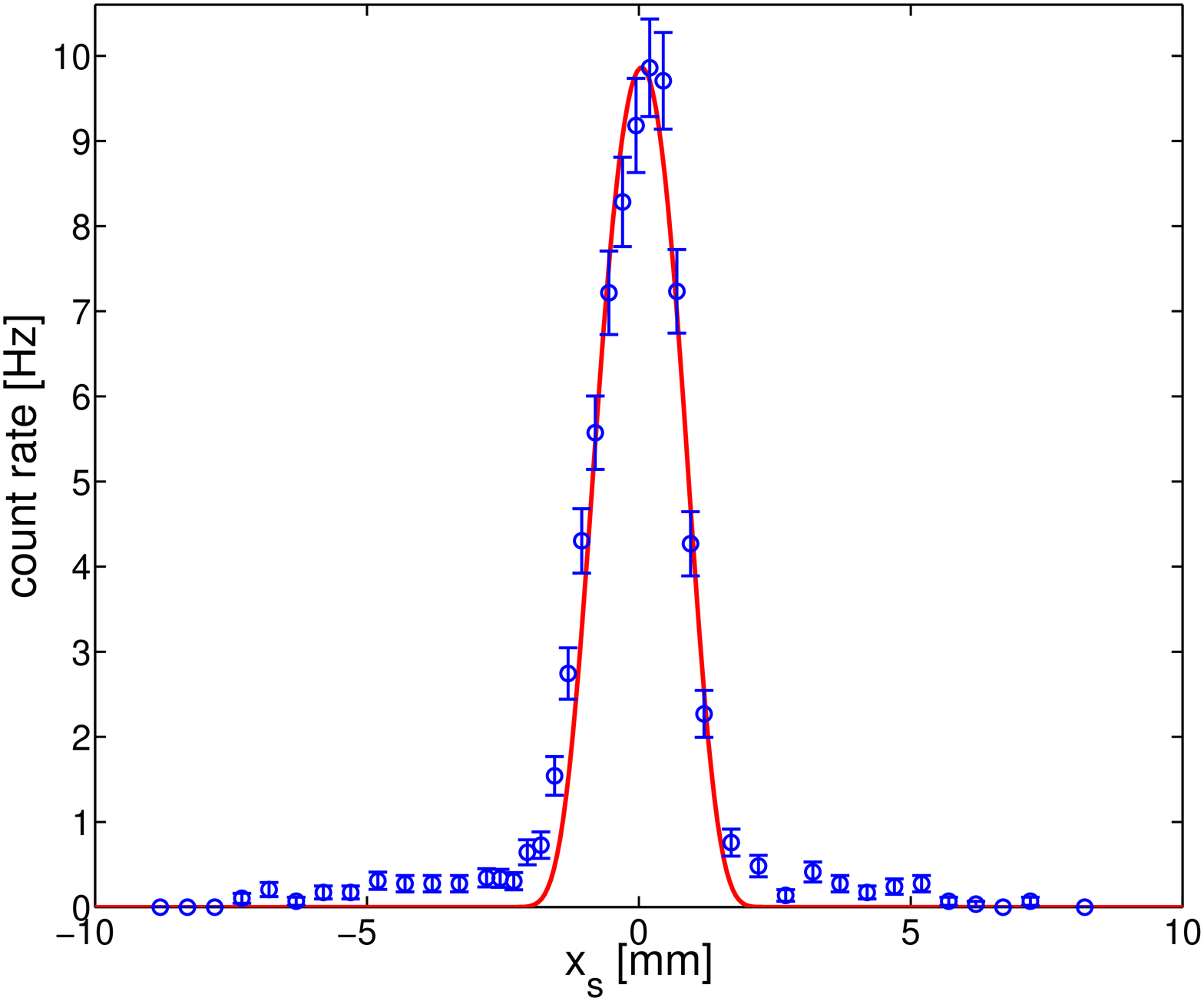}
\caption{Conditional distribution}
\label{fig:ConditionalCompensetion}
\end{subfigure}
\caption{Theoretical and experimental TPA distributions obtained in the configuration of anisotropy compensation.}
\label{fig:twoCrysCompensation}
\end{figure}

We then rotated the second crystal by $180^\circ$ in order to pass to the anisotropy compensation configuration. The optic axis of the second crystal was then at $-29.03^{\circ}$ and the phase matching was still fulfilled. In this case the TPA of the second crystal is bent oppositely with respect to that of the first one. The two contributions are clearly seen in Fig. \ref{fig:TPACompensation}. As expected, the transverse distributions are symmetric but still show interference.

\begin{figure}[htb]
\centering
\begin{subfigure}[b]{0.32\textwidth}
\includegraphics[width=\textwidth]{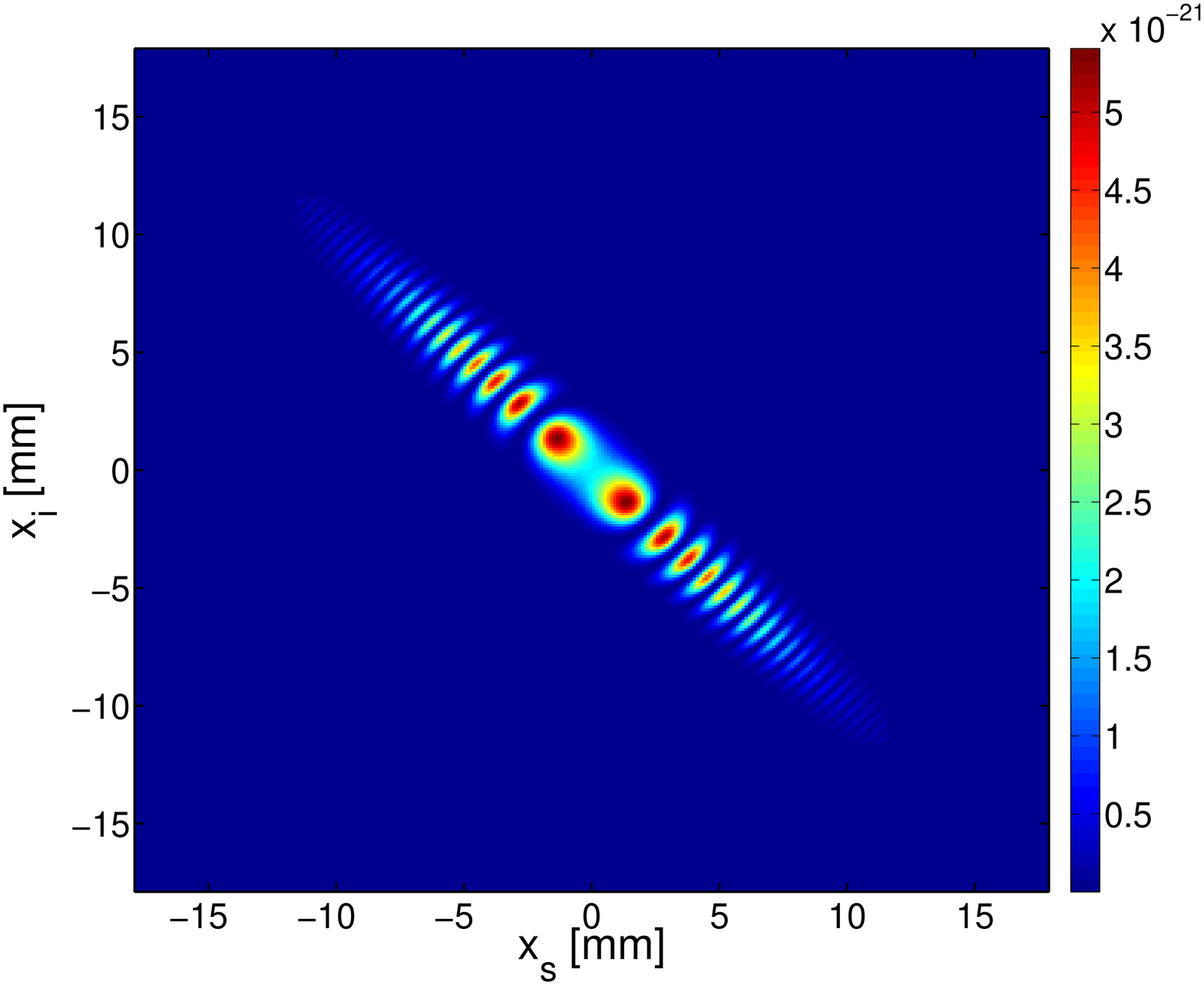}
\caption{TPA}
\label{fig:TPANoAnisotropy}
\end{subfigure}
\begin{subfigure}[b]{0.32\textwidth}
\includegraphics[width=\textwidth]{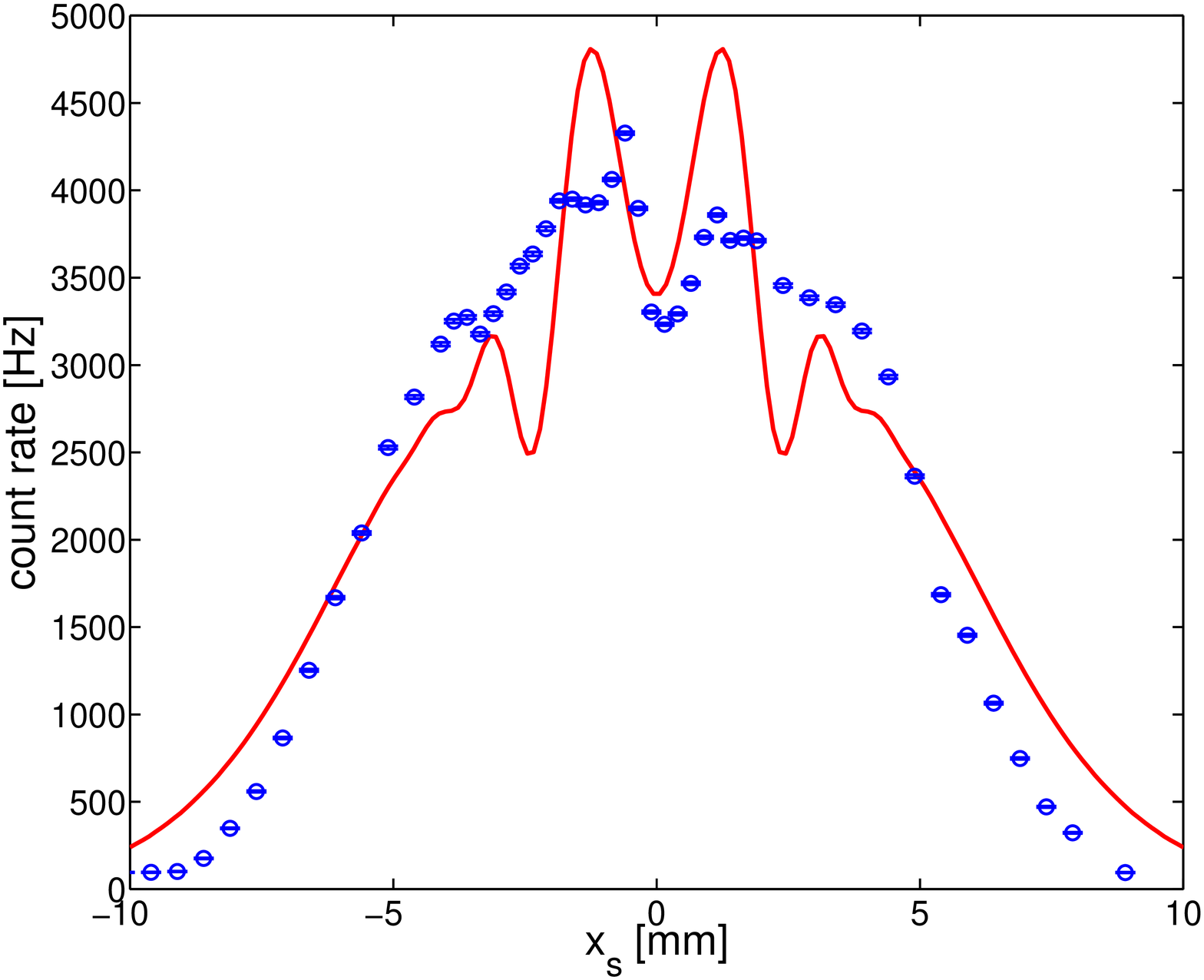}
\caption{Unconditional distribution}
\label{fig:UnconditionalNoAnisotropy}
\end{subfigure}
\begin{subfigure}[b]{0.32\textwidth}
\includegraphics[width=\textwidth]{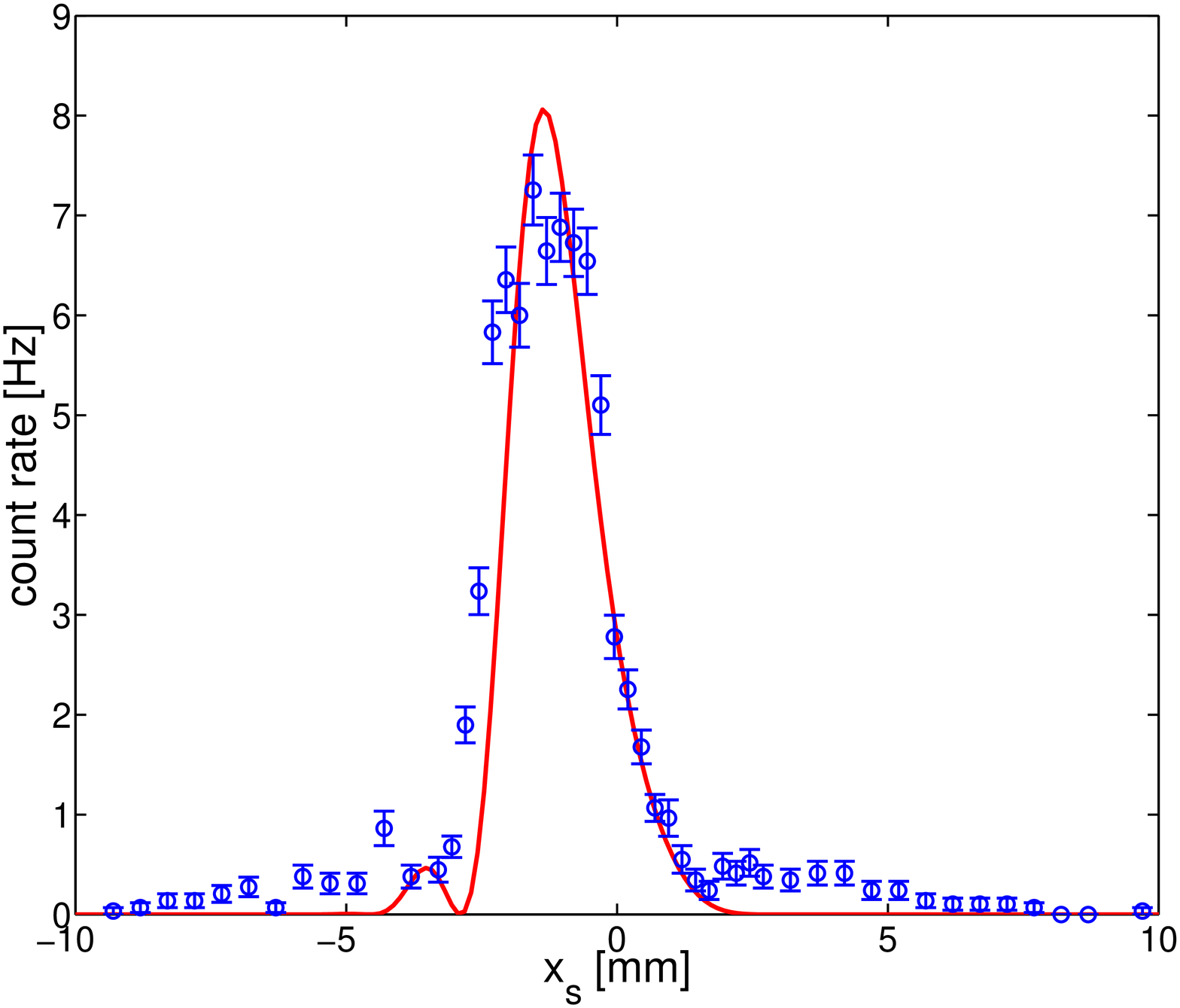}
\caption{Conditional distribution}
\label{fig:ConditionalNoAnisotropy}
\end{subfigure}
\caption{Theoretical and experimental TPA distributions for the `weak anisotropy' configuration.}
\label{fig:twoCrysNoAnisotropy}
\end{figure}

For the last measurement we rotated the crystals and the polarisation of the pump by $90^\circ$ in order to measure the distributions perpendicular to the principal plane, i.e., in the weak anisotropy configuration. Figure \ref{fig:twoCrysNoAnisotropy} shows the experimental data and the theoretical simulations.
Although in this configuration the effect of anisotropy is more subtle, both conditional and unconditional distributions also indicate it by a certain change in the width with respect to the case where anisotropy is absent.

As already mentioned, an important task is selecting a single Schmidt mode of two-photon light and this task is much simplified if the first (strongest) mode is Gaussian. The Schmidt decomposition for type-I degenerate collinear two-photon state is
\begin{equation}
|\psi\rangle=\sum_{n=0}^{\infty}\sqrt{\lambda_n}\phi_n^{(s)}\phi_n^{(i)},
\end{equation}
where $\phi_n^{(s)}$ and $\phi_n^{(i)}$ are the Schmidt modes for signal and idler photons (note that they are identical in our case). The Schmidt coefficient $\lambda_n$ gives the weight of a given mode $\phi_n^{(s,i)}$. Since the fundamental Gaussian is the most convenient mode to work with, it is logical to aim for a high corresponding Schmidt coefficient.

\begin{figure}[htb]
\centering
\begin{subfigure}[h]{0.32\textwidth}
\includegraphics[width=\textwidth]{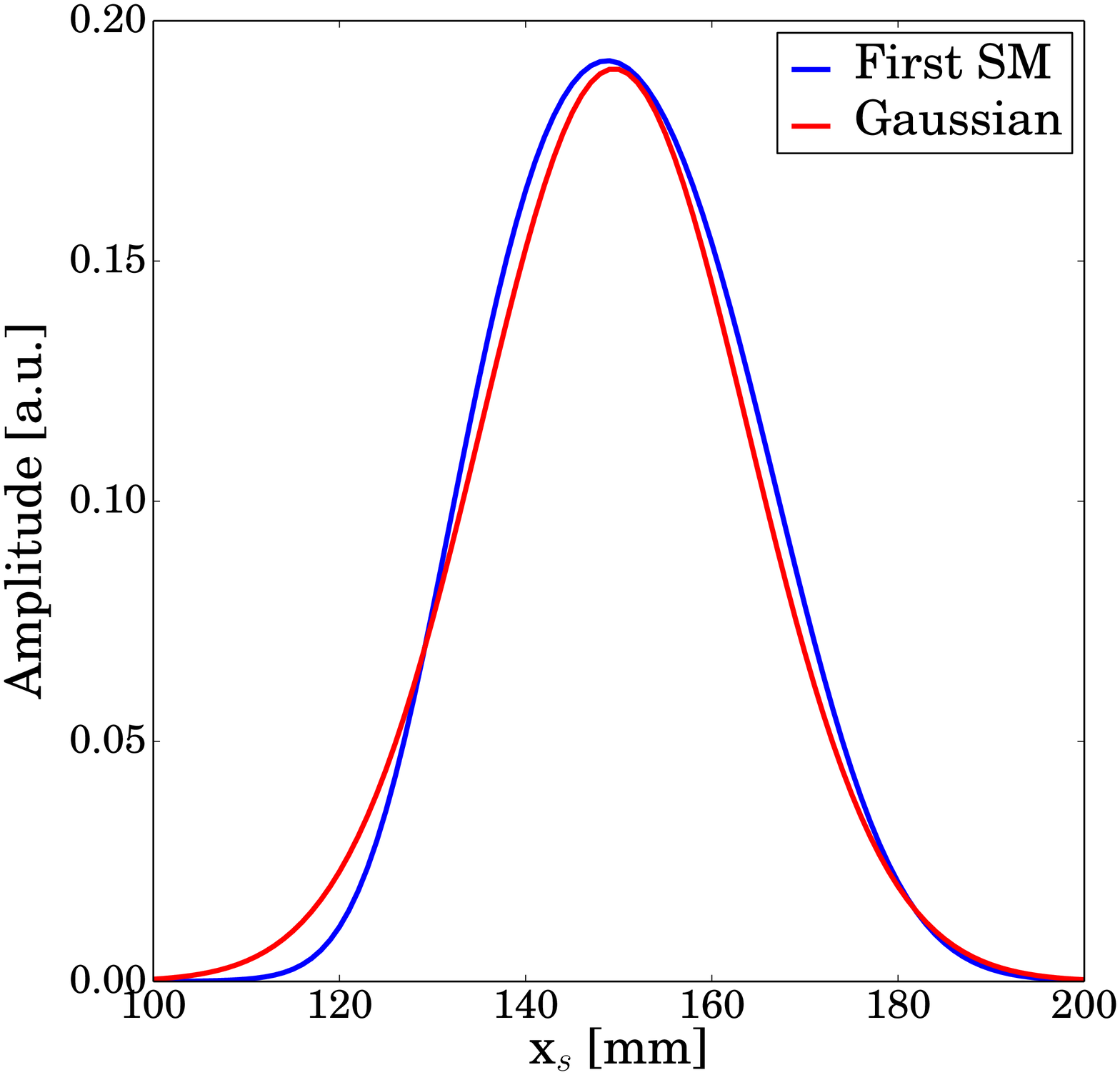}
\caption{One Crystal}
\label{fig:smonecrystal}
\end{subfigure}
\begin{subfigure}[h]{0.32\textwidth}
\includegraphics[width=\textwidth]{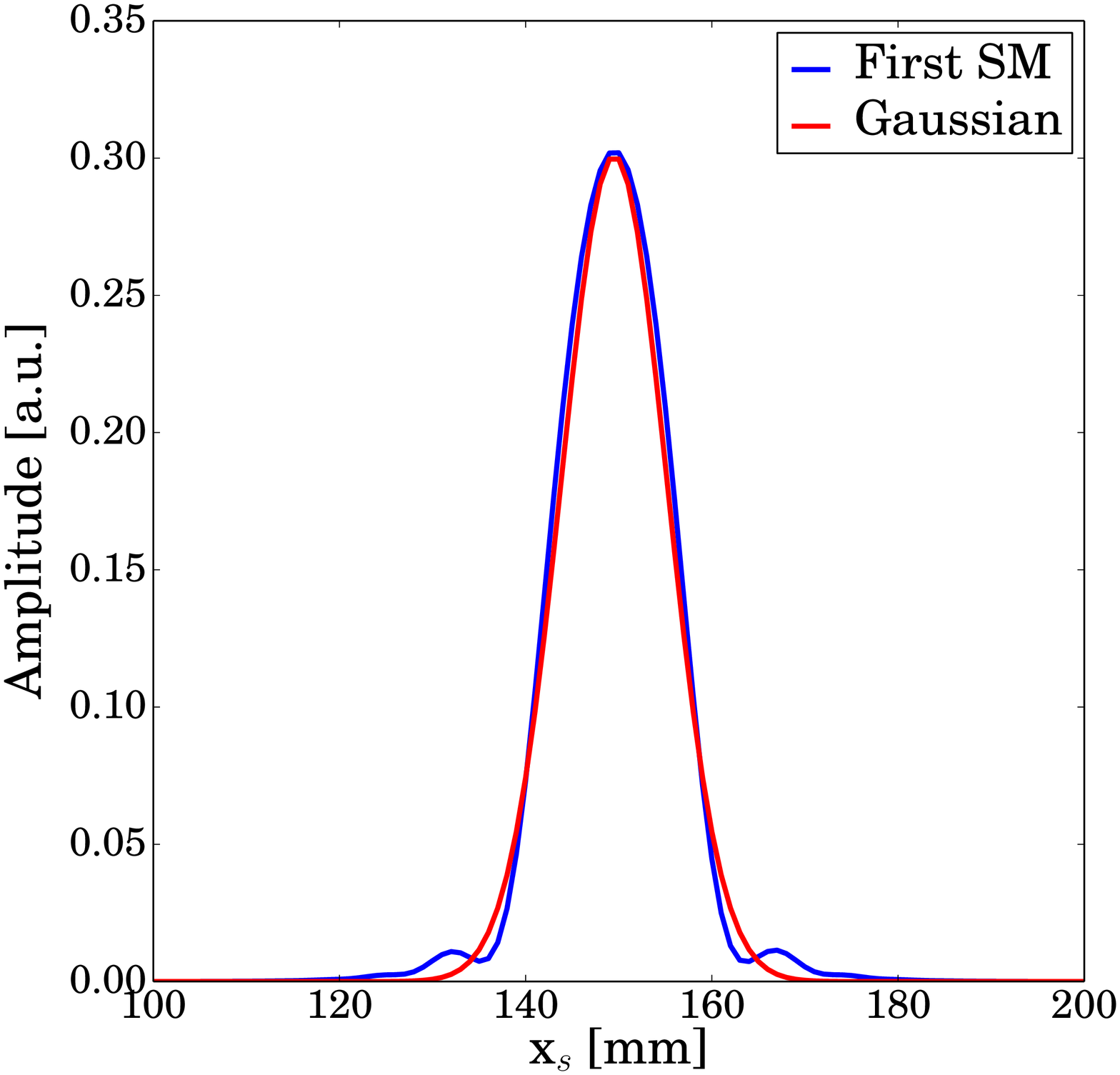}
\caption{Two crystals compensation}
\label{fig:smcompensation}
\end{subfigure}
\caption{a) First Schmidt mode for one crystal of 2 mm compared to a gaussian fit. b) First Schmidt mode for two crystals of 1 mm in the compensation configuration.}
\label{fig:sm}
\end{figure}

Using the TPAs calculated above, we numerically performed the Schmidt decomposition for the case of an individual 2mm crystal and also for the anisotropy compensation case using two 1mm crystals. In the first case the dominant Schmidt mode $\phi_0^{(s,i)}$ has a Gaussian shape (the overlap integral is 0.997) and a weight $\lambda_0= 0.094$ (Fig. \ref{fig:smonecrystal}). In the second case the dominant mode is still Gaussian with even a higher overlap integral (0.999) and a much larger weight $\lambda_0 = 0.15$ (Fig. \ref{fig:smcompensation}). Especially for bright squeezed vacuum applications where tight pump waists and long (or consecutive) crystals are a necessity, this is an important result.

\section{Conclusion}
We have studied the TPA for two-photon light emitted via PDC from two crystals oriented in such a way that the transverse walk-off is compensated. The compensation is achieved due to the opposite tilt of the optic axes in the two crystals and therefore opposite directions of the walk-off. Both the theoretical calculation and the experimental results show that, in contrast to the case of a single crystal or two crystals with the optic axes oriented parallel (non-compensating configuration), the TPA becomes closer to Gaussian and the angular distributions of both intensity and coincidences become symmetric.

Importantly, compensation of the anisotropy leads to a TPA whose Schmidt decomposition is much closer to that given by the double-Gaussian model than in the non-compensated case. The first Schmidt mode is very close to Gaussian and hence can be filtered out using a single-mode optical fibre. This is especially important at high-gain PDC, where bright squeezed vacuum is produced. The existence of the interference fringes adds non-Gaussian features to the TPA; however, by minimising the distance between the two crystals it is possible to have a maximum at the centre. In this situation the Schmidt decomposition feature a first Schmidt mode close to Gaussian.

This work was supported by ERA-Net.RUS (project Nanoquint).

\end{document}